\documentclass[twocolumn,showpacs,prl,preprintnumbers,floatfix,amsmath,amssymb]{revtex4}
\usepackage{graphicx}

\usepackage{dcolumn}
\begin{document}
\title{Orbital-Peierls State in NaTiSi$_2$O$_6$}
\author{Jasper van Wezel$^1$ and Jeroen van den Brink$^{1,2}$}
\affiliation{$^1$Institute-Lorentz for Theoretical Physics, Universiteit Leiden,\\
P.O.Box 9506, 2300 RA Leiden, The Netherlands\\
$^2$Institute for Molecules and Materials, Radboud Universiteit Nijmegen,\\ 
P.O. Box 9010, 6500 GL Nijmegen, The Netherlands}

\begin{abstract}
Does the quasi one-dimensional titanium pyroxene NaTiSi$_2$O$_6$ exhibit the novel {\it orbital-Peierls} state? We calculate its groundstate properties by three methods: Monte Carlo simulations, a spin-orbital decoupling scheme and a mapping onto a classical model. The results show univocally that for the spin and orbital ordering to occur at the same temperature --an experimental observation-- the crystal field needs to be small and  the orbitals are active. We also find that quantum fluctuations in the spin-orbital sector drive the transition, explaining why canonical bandstructure methods fail to find it. The conclusion that NaTiSi$_2$O$_6$ shows an orbital-Peierls transition is therefore inevitable.
\end{abstract}

 \maketitle

{\it Introduction.}
Transition-metal compounds display a dazzling collection of physical properties, spanning the range from colossal magneto-resistance in manganites, the spin-Peierls effect in quasi one-dimensional systems to high $T_{\rm c}$ superconductivity in cuprates~\cite{Tokura00,Imada98}. In these materials the interactions between electrons can be so strong that they destroy the Fermi surface. A Mott insulator is the canonical example. In this situation electrons become localized and they regain their local spin, charge and orbital degrees of freedom. The orbital degrees of freedom play a special role here since they can couple both to the lattice, via the Jahn-Teller effect, and to the spins, via superexchange interactions~\cite{Kugel82}. The interplay between all these degrees of freedom is the apparent root of the observed wealth in symmetry broken groundstates in e.g. doped manganites and even the driving force behind multiply ordered systems, such as e.g. multiferroics~\cite{Kimura03,Efremov04}.

Recently a novel type of groundstate, the {\it orbital-Peierls state}, was proposed to exist in several compounds. This state is expected to appear in one-dimensional systems where orbital degrees of freedom lead to the dimerization of the atoms that make up the chain, in analogy to the {\it spin-Peierls state}, where the dimerization is driven by the electron spins. As the orbital degrees of freedom couple to both the lattice and spin, the orbital-Peierls state is expected to reveal itself experimentally through orbital, spin and lattice anomalies. The search for this state, specifically in vanadates and titanates, motivated a number of experimental and theoretical studies~\cite{Ulrich03,Horsch03,Fang04,Ishihara05,Isobe02,Konstantinovic04,Hikihara04,Popovic04,Popovic05,Wezel05,Bersier,Streltsov05} and the findings became the subject of intense debate. The first candidate with possibly an orbital-Peierls groundstate is the perovskite vanadate YVO$_3$. However, the evidence in favor~\cite{Ulrich03,Horsch03} is opposed by extended band-structure calculations that indicate that the compound is electronically three-dimensional~\cite{Fang04}. In this letter we focus on the other possible candidate~\cite{footnote1}, the pyroxene compound NaTiSi$_2$O$_6$. Studies of its microscopic spin-orbital Hamiltonian have shown that indeed this novel state can form, but {\it ab initio} LDA bandstructure and ionic crystal field calculations contradict this claim~\cite{Isobe02,Konstantinovic04,Hikihara04,Popovic04,Popovic05,Wezel05,Bersier,Streltsov05}. Here we present a unified picture and show that indeed it exhibits orbital-Peierls ordering.

{\it The titanium pyroxene.} 
The distinguishing feature of the pyroxene compound NaTiSi$_2$O$_6$ is the presence of quasi one-dimensional arrays of edge sharing TiO$_6$ octahedra. Besides its spin (s=1/2), each Ti$^{3+}$ ion has an additional orbital degree of freedom. The system shows a dramatic drop in the magnetic susceptibility below $T_{OO} = 210~K$~\cite{Isobe02}, while at the same temperature a structural change is observed. Raman scattering reveals that this transition is accompanied by a phonon frequency shift and broadening~\cite{Konstantinovic04,Popovic05}. It was immediately suggested that this apparent connection between lattice and spin degrees of freedom is driven by the ordering of titanium $t_{2g}$ orbitals. Starting from a tight binding Hubbard model a microscopic spin-orbital model was derived in support of this picture and it was shown explicitly that the spin-orbital model can have an orbital-Peierls groundstate at low temperatures~\cite{Konstantinovic04,Hikihara04}.

However, also two pieces of evidence disputing this claim have appeared. Ionic crystal field calculations suggest that the $t_{2_g}$ orbital splitting is substantial in this compound~\cite{Bersier}. Since the microscopic models are based on the assumption that the orbitals are (nearly) degenerate, these results seem to strike a serious blow to the proposed orbital-Peierls picture. Besides this, density functional calculations by Popovi\'c {\it et al.} indicate that a {{\it Haldane spin-one chain} forms at low temperature, preempting the orbital-Peierls transition~\cite{Popovic04}. Recently this result was challenged by LDA+U bandstructure calculations, where electron correlation effects are partially incorporated on a mean field level~\cite{Streltsov05}.

In this Letter we will show that we can account for all of these results, by an analysis that is based on the microscopic model that one of us originally proposed in Ref.~\cite{Konstantinovic04}. Both numerical Monte Carlo simulations and an analytical spin-orbital decoupling for the microscopic model show that even if the crystal field splitting is comparable to the size of the inter-site exchange coupling, the orbital-Peierls transition still occurs. Thus for the spin and orbital ordering to occur at the same temperature --as it does experimentally-- the crystal field should be smaller than or at most comparable to the inter-site exchange energy. This places a strict upper bound on the crystal-field splitting.
Furthermore we find that quantum fluctuations in the spin and orbital sector actually drive the orbital-Peierls transition. Precisely these correlations are not taken into account in canonical band-structure calculations. Neglecting them in our microscopic Hamiltonian actually reduces it to a solvable classical model. An exact transfer matrix computation shows that without the quantum fluctuations indeed alternating ferromagnetic correlations between neighboring spins appear, as in bandstructure calculations of Ref.~\cite{Popovic04}. This explains the failure of this method to find the orbital-Peierls groundstate. Judging from these results the conclusion that NaTiSi$_2$O$_6$ exhibits an orbital-Peierls transition is inevitable.

{\it Microscopic Model.} 
Each Ti$^{3+}$ ion in the edge-sharing TiO$_6$ octahedra has a single electron in one of the three $t_{2g}$ orbitals denoted by $|xy\rangle $, $|yz\rangle $ and $|zx\rangle$. The $|zx\rangle$ orbital turns out to be inert, which renders it higher in energy and it is thus neglected. The Coulomb interaction between electrons on the same titanium atom is so large that the exchange interactions can be determined by a second order perturbation expansion in the electron hopping parameter. The leading order spin-orbital Hamiltonian is~\cite{Konstantinovic04}:
\begin{equation}
H_{ST} = 4J \sum_{\langle ij \rangle } {\bf S}_i \cdot {\bf S}_j \left[ T_i^z T_j^z + \frac{(-1)^i}{2} \left(T_i^z+T_j^z\right)+ \frac{1}{4} \right]
\label{H_ST}
\end{equation}
where we use spin operators ${\bf S}$ and orbital operators $T$ (where $T_z=\pm \frac{1}{2}$ and the plus (minus) sign correspond to an occupied $|xy\rangle$ ($|yz\rangle$) orbital). Nearest neighboring lattice sites are denoted by $i$ and $j$ and $J$ denotes the exchange integral. Here we neglected the small contributions due to the Hund's rule exchange, but we checked for each case that our final results do not depend on this approximation. We also include two orbital-only contributions to the Hamiltonian, the crystal field splitting, lifting the degeneracy of the two $t_{2g}$ orbitals, parameterized by $J_{\text{CF}}$, and a term due to the superexchange~\cite{Hikihara04}:
\begin{equation}
H_T = J \sum_{\langle ij \rangle } T_i^z T_j^z + J_{\text{CF}} \sum_i T_i^z,
\label{H_T} 
\end{equation}

\begin{figure}
\includegraphics[width=1.0\columnwidth]{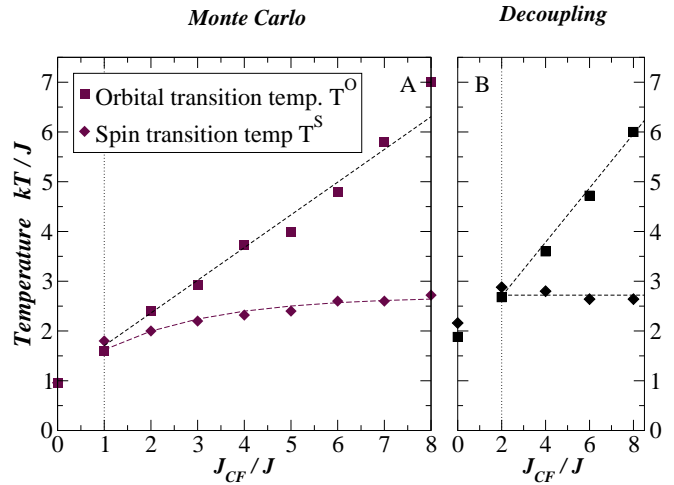}
\caption{Left: orbital and spin transition temperatures versus crystal field splitting determined from the Monte Carlo calculations. In the region left of the dotted line the spin and ordering temperature coincide. Right: transition temperatures versus crystal field splitting from the spin-orbital decoupling calculations.}
\end{figure}

{\it Methods.} 
We study the model given by equations~(\ref{H_ST}) and (\ref{H_T}) by three different methods. First we performed a Monte Carlo simulation of the Hamiltonian for a system of 100 sites. In this simulation we added an inter-chain coupling in a mean field fashion, in order to account for the three dimensionality of the system:
\begin{equation}
H = H_{ST} +H_T + J_{\text{inter}} T_{tot}^z \  {\rm and}  \   J_{\text{inter}}=\frac{4z J'}{N} \left< T_{tot}^z \right>.
\label{H}
\end{equation}
Here $J_{\text{inter}}$ is the effective mean field inter-chain coupling, $z$ is the number of neighbor sites and $J'$ the bare inter-chain coupling strength. We take $J'=J/10$ in the present calculations. It is worthwhile to note that one can use a classical Monte Caro simulation for this Hamiltonian, rather than a quantum Monte Carlo algorithm. This is justified by a particular symmetry in the spin-orbital model: the staircase structure of the titanium chains combined with the orientation of the $t_{2_g}$ orbitals does not allow any spin states other than single spins or dimers.

Within the Monte Carlo simulation we determine the magnetic (spin) and orbital susceptibilities, $\chi^S$ and $\chi^O$, as well as the nearest neighbor correlators for spins and orbitals: $C^S$ and $C^O$ respectively. From the temperature dependence of the spin/orbital susceptibility we can then also deduce the transition temperatures below which spin and orbital ordering are found ($T^S$ and $T^O$). These quantities we studied as a function of both $J$ and $J_{\text{CF}}$.

In the second method we employ a decoupling of the spin and orbital degrees of freedom of the Hamiltonian~(\ref{H_ST}). We decouple $H_{ST}$ by introducing static orbital and spin fields:
\begin{align}
s+(-1)^i \delta s & \equiv \left< {\bf S}_i \cdot {\bf S}_j \right> \nonumber \\
t+(-1)^i \delta t & \equiv \left< \frac{1}{4} + T_i^z T_j^z + \frac{(-1)^i}{2} \left(T_i^z+T_j^z\right) \right>,
\end{align}
so that the decoupled part of $H_{ST}  \simeq H_{\text{de}}^S + H_{\text{de}}^O $ with
\begin{align}
H_{\text{de}}^S &= 4J\sum_{\langle i,j\rangle } \left(t+(-1)^i \delta t\right){\bf S}_i \cdot {\bf S}_j \nonumber\\
H_{\text{de}}^O &= \sum_{\langle i,j\rangle } \left[ 4J\left(s+(-1)^i \delta s\right) + J \right] T_i^zT_j^z \nonumber\\
& ~ + \sum_i \left( 4J ~\delta s + J_{\text{CF}} \right) T_i^z.
\label{H_MF}
\end{align}

\begin{figure}
\includegraphics[width=1.0\columnwidth]{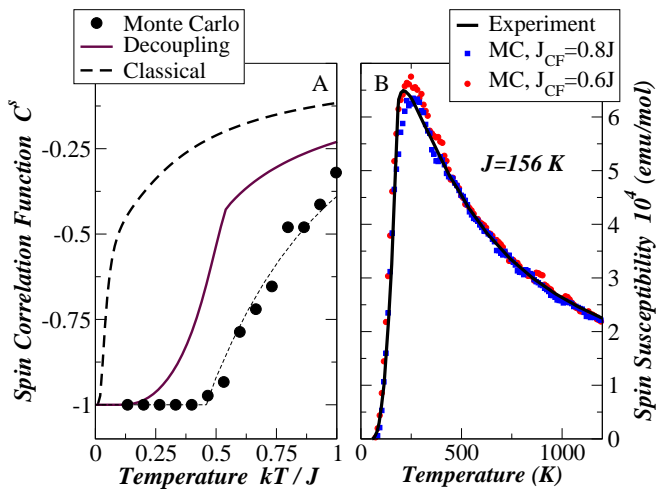}
\caption{Left: nearest neighbor spin correlation function as calculated from the Monte Carlo simulation, the spin-orbital decoupling scheme, and the classical  model. The correlation functions are normalized to their zero temperature values (-3/8, -1/4 and -1/8, respectively). In the first two cases spin valence bonds are formed, in the last case small residual interactions due to Hund's rule coupling will cause the formation of a chain of alternating ferro and antiferro bonds, see Fig.~3.
Right: fit of the Monte Carlo results to the experimental spin susceptibility.}
\end{figure}

We focus first on the orbital part of this decoupled Hamiltonian. Since the only operator appearing in $H_{\text{de}}^O$ is $T^z$, it is effectively a classical Hamiltonian, closely related to the Ising model. This makes it possible to construct a transfer matrix that makes it possible to obtain an exact expression for the partition function. From there we can obtain any other thermodynamic quantity related to the orbital physics of $H_{\text{de}}^O$.
The spin part of the decoupled Hamiltonian~(\ref{H_MF}) cannot be diagonalized exactly as it stands. Therefore, following a canonical approach to the spin-Peierls Hamiltonian~\cite{Cross79}, we consider only the XY part of the spins. Then we can use the Jordan-Wigner transformations to diagonalize the remaining part of $H_{\text{de}}^S$, and find all spin related thermodynamic quantities as modeled by $H_{\text{de}}^O$.

Finally, we have considered the full Hamiltonian~(\ref{H}) without quantum spin fluctuations, i.e. taking only into account the $z$ components of the spin operators. The motivation for considering this limit is that in bandstructure calculations quantum fluctuations of this kind are effectively discarded. 
Since this Hamiltonian is classical, we can construct the transfer matrix that will allow us to find the partition function and thus all thermodynamic quantities of this system.

\begin{figure}
\centerline{\includegraphics[width=0.6\columnwidth]{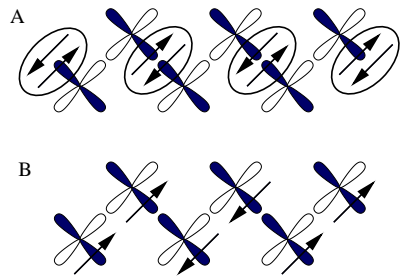}}
\caption{\label{pic} A sketch of the zero temperature states. The lobes of the orbitals that lie in the plane of the titanium chain are depicted. The dark lobes are occupied orbitals. Top: orbital-Peierls state with spin valence bonds. Bottom: alternating ferro and antiferro spin bonds as found in the classical spin model and density functional studies.}
\end{figure}

{\it Results.} 
In Fig. 1 the spin and orbital ordering temperatures are shown as a function of the crystal field splitting $J_{CF}$, as deduced from their respective susceptibilities in the Monte Carlo calculations. 
The transition temperatures for magnetic ordering coincide with the ones for orbital ordering, as long as the crystal field splitting is not too large, i.e., when $J_{CF}/J \lesssim 1$. The calculations show that the orbital and magnetic ordering occur at different temperatures only for a crystal field that is larger than the exchange coupling, i.e. when $J_{CF}/J \gtrsim 1$. In that case $T^O$ is proportional to $J_{CF}$, whereas $T^S$ tends to a constant. The same conclusion we find from decoupling calculations, shown in the right pannel of Fig.~1.

We now invert the argument above. Experimentally the drop in spin susceptibility and the structural changes that are indicative of orbital ordering occur both at $T_{OO}= 210 \ K$. As experiment shows that these temperatures coincide, the splitting of $t_{2_g}$ levels can thus not be larger than the superexchange coupling of neighboring spins: evidently titanium pyroxene is in the regime where $J_{CF}/J < 1$.
This conclusion can be further quantified. In Fig.~2, right pannel, we compare the intrinsic experimental spin susceptibility from Ref.~\cite{Isobe02} and our Monte Carlo results. From fitting the experimental susceptibility we find $J_{CF}/J = 0.7 \pm 0.1$, $J= 156 K$ and a gyromagnetic moment $g=1.85$. 

We also calculated the nearest neighbor spin and orbital correlation function $C^S$/$C^O$ as a function of temperature with all three methods. The results for the spin correlator are shown in figure Fig.~2, left pannel. The corresponding orbital correlators $C^O$ indicate that in all cases there is a uniform and parallel ordering of the orbitals at low temperatures and are not shown. The correlators in Fig.~2A are, for clarity, normalized to their zero temperature value. Clearly spins tend to align antiparallel on neighboring bonds and form a valence bond solid, as shown in Fig.~3A. The Monte Carlo results show that this tendency is very strong and that even above the ordering temperature spin dimer correlations are substantial. It is interesting to note that when the spin and orbital fluctuations are decoupled, the spin correlation function $C^S$ is appreciably reduced at finite temperatures. Thus the simultaneous fluctuations of spin and orbital degrees of freedom actually strengthen the spin valence bond solid. This seems counter intuitive, but has a straightforward physical cause. A spin flip in the valence bond state transforms a singlet dimer into a triplet, which is a strong ferromagnetic bond that causes an large reduction of the average antiferromagnetic spin correlations. Suppose, however, that the flipped spin can at the same time change orbital. After such a combined spin-orbital excitation, the spins that originally formed a firm singlet valence bond, are effectively not interacting anymore. They have a vanishing contribution to the nearest neighbor spin correlation function because of their respective orbital configurations and thus are less damaging to the magnetic ordering than fluctuations that exclusively involve the spin degree of freedom. 

The importance of fluctuations becomes even more striking when we compare the classical to the quantum results. When we only keep in the Hamiltonian $H_{ST}$ only the $z$ projection of the spin operators, as is the case in the density functional calculations, the spin correlation function $C^S$ is drastically reduced at any finite temperature, see Fig.~2A. Only at the lowest temperatures long range spin/orbital ordering sets in, which indicates that the system is strongly fluctuating due to the presence of many low energy states that are quasi degenerate. In this case we expect that the system is very susceptible to any residual spin interaction. Indeed, we find that if we include a finite Hund's rule coupling, which leads to a very small ferromagnetic interaction between neighboring spins that are in different orbitals, the spins immediately order ferromagnetically along the $\pi$-like bonds, whereas the spins on the $\sigma$-like bonds, with orbitals pointing towards each other, order antiferromagnetically, see Fig.~3B.
In this case nearest neighbor correlation function $C^S$, which is now the sum of antiferro and ferro correlations, goes to zero at $T=0$. The resulting ground state spin and orbital configuration (Fig.~3B) exactly matches the one found by Popovi\'c et al. in Ref. \cite{Popovic04} on the basis of bandstructure calculations. We find, however, that the ferro bonds are weak and the antiferro ones are strong, in accordance with recent LDA+U calculations~\cite{Streltsov05}. Therefore even in this classical limit the system cannot be regarded to be a spin one chain.

{\it Conclusions}. 
Results from Monte Carlo simulations, from a spin-orbital decoupling and a classical approximation scheme of a microscopic Hamiltonian which incorporates both spins and orbitals of the titanium chains in NaTiSi$_2$O$_6$ place a firm upper bound on the size of the crystal field splitting. This proves that in this system the orbital degrees of freedom are active at low energies and temperatures. The interacting spin-orbital Hamiltonian describes the experimental magnetic suscpetibility very well. The calculated correlation functions show that the orbital-Peierls state is realized below $T_{OO}=210 K$ and that quantum fluctuations in both the spin and orbital sector are essential for the stabilization of this spin-orbital solid.

{\it Acknowledgments.}
We thank Milan Konstantinovic, Jan Zaanen and Daniel Khomskii for stimulating discussions. 
This work is supported by FOM and the Dutch Science Foundation.

\end{document}